\documentclass{article}

\usepackage[utf8]{inputenc}
\usepackage[english]{babel}

\usepackage{natbib}
\setcitestyle{numbers,open={[},close={]},comma}

\usepackage{graphicx}
\usepackage{hyperref}
\hypersetup{colorlinks=true, allcolors=blue}
\usepackage{subcaption}

\usepackage{amsmath}
\usepackage{amsfonts}
\usepackage{amssymb}
\usepackage{amsthm}
\usepackage{mathrsfs}

\usepackage{lineno}

\usepackage{authblk}
\usepackage{booktabs}
\usepackage[labelfont=bf,labelsep=period]{caption}
\usepackage[flushleft]{threeparttable}

\newenvironment{revs}
{\begin{color}{black} \ignorespaces} %set "black" when you want the submission ready manuscript.
{\end{color}}

\usepackage[table,x11names,dvipsnames,table]{xcolor}

\definecolor{lightgray}{gray}{0.95}

\usepackage{subcaption}

\usepackage{commado,filesdo}

\begin{document}

\title{Covid19:
unless one gets everyone to act, policies may be ineffective or even backfire\thanks{%
We thank Alberto Dalmazzo, Matthew Jackson and Alessia Melegaro for very helpful comments. We gratefully acknowledge
funding from the Italian Ministry of Education Progetti di Rilevante Interesse
Nazionale (PRIN) grant 2017ELHNNJ.}}

\author[a]{Alessio Muscillo}
\author[a,b]{Paolo Pin}
\author[a,c]{Tiziano Razzolini}

\affil[a]{Department of Economics and Statistics, Universit\`a di Siena, Italy}
\affil[b]{IGIER \& BIDSA, Bocconi, Milan, Italy}
\affil[c]{IZA \& LABOR}

\date{\today}

\maketitle

\begin{abstract}
The diffusion of Covid-19 \begin{revs} has called\end{revs} governments and public health authorities to interventions \begin{revs} aiming at limiting   new infections and containing\end{revs} the expected number of critical cases and deaths. Most of these measures rely
%\begin{revs} have relied \end{revs} 
on the compliance of people, who are
%\begin{revs} have been \end{revs} 
asked to reduce their social contacts to a minimum.
In this note we argue that individuals' adherence to prescriptions and reduction of social activity 
%are not enough if done in a way that is not strong and homogeneous across the population.
may not be
%\begin{revs} have been \end{revs} 
efficacious if not implemented robustly on all social groups, especially on those characterized by intense mixing patterns.
Actually, it is possible that, if those who have many contacts \begin{revs} have reduced\end{revs} them proportionally less than those who have few, then the effect of a policy could \begin{revs} have backfired\end{revs}: the disease \begin{revs} has taken\end{revs} more time to die out, up to the point that it \begin{revs} has become\end{revs} endemic.
In a nutshell, unless one gets everyone to act, and specifically those who have more contacts, a policy may even be counterproductive.
\end{abstract}

%\linenumbers

% \section*{Abstract}

% The diffusion of Covid-19 \begin{revs} has called\end{revs} governments and public health authorities to interventions \begin{revs} aiming at limiting   new infections and containing\end{revs} the expected number of critical cases and deaths. Most of these measures rely
% %\begin{revs} have relied \end{revs} 
% on the compliance of people, who are
% %\begin{revs} have been \end{revs} 
% asked to reduce their social contacts to a minimum.
% In this note we argue that individuals' adherence to prescriptions and reduction of social activity 
% %are not enough if done in a way that is not strong and homogeneous across the population.
% may not be
% %\begin{revs} have been \end{revs} 
% efficacious if not implemented robustly on all social groups, especially on those characterized by intense mixing patterns.
% Actually, it is possible that, if those who have many contacts \begin{revs} have reduced\end{revs} them proportionally less than those who have few, then the effect of a policy could \begin{revs} have backfired\end{revs}: the disease \begin{revs} has taken\end{revs} more time to die out, up to the point that it \begin{revs} has become\end{revs} endemic.
% In a nutshell, unless one gets everyone to act, and specifically those who have more contacts, a policy may even be counterproductive.

\section*{Introduction}

As social scientists, we use epidemic models to mimic the diffusion of opportunities and ideas in the society.
In this context, we are used to think of the effects of people's choices and actions on diffusion processes, like viral marketing campaigns or the launch of new technologies that increase online contacts. In general, our focus is on what happens when each individual in the society takes autonomous decisions that affect her socialization.
Looking at people's responses and decisions can be helpful to design policies against diseases and to understand how they affect the behavior of other members of the society.

Following the outbreak of the new Coronavirus, governments have faced the necessity 
%to take decisions 
to foster the limitation of social contacts.  
In most cases, initially the population have been asked to limit their contacts relying on the individual sense of responsibility 
(an extreme case was the initial approach in the United Kingdom, where isolation was intended only for people suspected to be infected or arrived from abroad, as stated by the \href{https://www.legislation.gov.uk/uksi/2020/129/contents/made}{Health Protection (Coronavirus) Regulations 2020} on March 10); while only at a later stage rigid temporary laws have been issued (like China on January 23 and 26, 2020, or Italy on March 4 and 11).
% {\color{red} There are still countries like Sweden, where on March 27 the Government banned only public gatherings of more than 50 people.}
However, independently of the nature of the restrictions, it is clear that not every individual responds in the same way to impositions and requests. 
Some people cut immediately all their social contacts, while others may only marginally reduce them.
The classic argument of \emph{revealed preferences} suggests that those who have more social relations will be less prone to limit them: their behavior reveals that they care more than others about social interactions, for personal taste or for professional reasons.
So, if we just ask people to reduce their contacts at a level they feel safe, everybody will trade off the expected risks with the benefit that they perceive from socialization. Thus, those who have many contacts every day will be proportionally less inclined to cut them, compared than those who have few.

\begin{revs}
Available data and public concern often focus on the number of contacts that people have, with the obvious implication that a reduction in the average number of contacts would correspond to a reduction in the infections.
In this note, we use an empirical analysis and a stylized model to show that, together with the average number of contacts, other measures of statistical dispersion -- i.e. squared number of contacts (roughly, variance) -- are also important to explain the variation in the number of infections.
\end{revs}

\section*{Empirical motivation}

\begin{revs}
We conduct an exploratory analysis using the public dataset provided by Belot and colleagues \cite{belot2020six}. This dataset contains, among other information, survey data on the individual number of contacts in the regions of six countries before and after the interview, occurred in the third week of April 2020.
The respondents to the survey were asked the number of their contacts before the outbreak of Covid-19 and in the last two weeks preceding the interview. 
% For each individual we have measured in each period  the total number of contacts occurred with any individual (children, adults and elderly people) for more than 15 minutes. 
We have then computed the average number of contacts and average squared number of contacts at the regional level.
% , respectively denoted by $\langle d \rangle$ and $\langle d^2 \rangle$ in Table \ref{tab:varcovid}.

We limit our analysis to the regions of Italy, South Korea and the United Kingdom, because of the large impact of the coronavirus in these countries and of the availability of data at a regional level.
For each region, we have collected the weekly number of cases and deaths before and after the survey interview 
%(i.e. third week of April) 
using the publicly available datasets listed in the Supporting Information. 

Italy and South Korea displayed a similar timing in the diffusion of coronavirus as well as in government interventions, as shown by the \emph{government response stringency index} developed by Hale and colleagues \cite{hale2020oxford}. 
For these two countries we have computed the weekly number of confirmed cases and deaths in the week from March 1 to March 8 for the pre-interview period and in the week from April 13 to April 20 for the post-interview period. 
Due to the later diffusion of coronavirus in the UK, the weekly number of confirmed cases and deaths is measured during the week from March 13 to March 20 and the week from April 24 to May 1.\footnote{In the UK, the indexes on the diffusion of Covid-19 are reported on a weekly basis, starting on Fridays.}

% Our main dependent variables are the regional variation in the number of confirmed cases and in the number of deaths in the pre- and post-interview period ($\Delta$Confirmed Cases, $\Delta$Deaths). 
% We consider the number of confirmed cases and deaths in a region, before and after the interview. Then, our main dependent variables are the pre- and post-interview variation of these numbers, that is, $\Delta$Confirmed Cases and $\Delta$Deaths.
We perform two ordinary least squares regressions where the dependent variables are the regional pre- and post-interview variation in the number of confirmed cases and deaths (respectively, $\Delta$Confirmed Cases and $\Delta$Deaths). 
The number of contacts that an individual has is denoted by $d$, so that the (regional) average number of contacts is denoted by $\langle d\rangle$ and the (regional) average squared number of contact is $\langle d^2\rangle$. The explanatory variables used in the regression are the pre- and post-interview variation: $\Delta\langle d \rangle$ and $\Delta\langle d^2 \rangle$.

The estimates are shown in Table \ref{tab:varcovid}. 
Columns (1) and (2) report the estimated effects on the variation in the number of confirmed cases while columns (3) and (4) the estimated effects on the variation in the number of deaths.
The variation in the number of contacts, $\Delta\langle d \rangle$, always displays a negative effect significantly different from zero on both dependent variables, as an obvious result of the endogenous reaction to the spread of coronavirus.\footnote{This is because causality in the real data goes in both directions, and the larger is the diffusion of the virus, the larger is the average reduction in contacts.}
The variation in the squared number of contacts, $\Delta\langle d^2 \rangle$, has a positive effect always statistically different from zero at 5\% significance level.
% Since the square of the average number of contacts is a component of the variance we have included its variation as a separate regressor. 
% The estimated effect of the variation in variances remain positive and statistically significant. 
The inclusion of the latter regressor always improves the adjusted $R^2$, thus indicating that this variables has a great explanatory power.\footnote{In the Supporting Information we also test the presence of non-linear effects including as a regressor the variation in the square of the average number of contacts.}

\renewcommand{\arraystretch}{1.3}
\begin{table}[h]
%\color{red}
\centering
\small
\addtolength{\tabcolsep}{5pt}
 \begin{threeparttable}
	\caption{\textbf{Variations in confirmed cases and deaths}}
	\label{tab:varcovid}
	%\rowcolors{1}{}{lightgray}
	\begin{tabular}{l|cc|cc}
		\hline\hline
		&    \multicolumn{2}{c}{$\Delta$ Confirmed Cases}  &   \multicolumn{2}{c}{$\Delta$ Deaths}  \\
		&        (1)   &     (2)   &        (3)   &        (4)    \\
		\hline
		$\Delta \langle d \rangle$    &     -69.228***&    -167.411***&     -25.549***&     -50.594***\\
		&    (21.289)   &    (51.019)   &     (8.300)   &    (14.582)   \\
		$\Delta \langle d^2\rangle$ &               &       0.382** &               &       0.097** \\
		&               &     (0.172)   &               &     (0.044)   \\
		Constant       &     213.419** &     163.936   &      92.514***&      79.892** \\
		&   (102.497)   &   (113.241)   &    (33.942)   &    (33.278)   \\
		N. obs.           &      48   &      48   &      48   &      48   \\
		Adj. $R^2$      &       0.100   &       0.231   &       0.158   &       0.245   \\
		\hline\hline	
\end{tabular}
\begin{tablenotes}
      \small
      \item Note: In columns 1 and 2 the dependent variable is the variation in the numbers of confirmed cases in a region. In columns 3 and 4 the dependent variable is the variation in the number of deaths in each region. $\Delta\langle d\rangle$ is the variation in the average number of contacts in each region. $\Delta \langle d^2\rangle$ is the variation in the average squared number of contacts in each region. Robust standard errors in parentheses. *** significant at 1\%, ** significant at 5\%, * significant at 10\%.
    \end{tablenotes}
  \end{threeparttable}
\end{table}

\end{revs}

\section*{SI--type model}

\begin{revs} In this section, we build a stylized model to\end{revs} argue that when a disease spreads in a population with heterogeneous intensity of meetings -- a so-called \emph{complex network} -- if the individuals who meet many people exhibit high resistance against isolation policies, such policies may not only turn out to be ineffective, but can even be detrimental.
Imagine a social-distancing policy that asks people to limit their contacts to reduce the diffusion of a disease. This generates a new reduced social network that is smaller but denser.
The main unintended negative consequence of the policy could be that even if the disease was eventually going to die out in the original social network, it becomes endemic in the new network instead. Paradoxically, as long as the policy is in force, the disease will be kept alive.

The intuition behind the phenomenon is simple to grasp, and we leave the details of a parsimonious susceptible-infected-susceptible (SIS) model in the Supporting Information.
Consider the social network of a society, where some people have few links and others have many. Consider, also, a disease spreading via these contacts. 
Whether the disease will be \emph{endemic} or not turns out to depend on the interplay between the features of the disease itself and the statistical properties of the social network through which it is spreading.
The disease is concisely described by the transmission rate, $\beta$, and the recovery rate, $\delta$. 
The characteristics of the social network are captured by the number of contacts that one has $d$. In particular, by its average across people, denoted $\langle d \rangle$, and by the expected square of this number, $\langle d^2 \rangle$.

Whether the disease dies out or remains endemic depends on the relationship between two quantities:
$$
\lambda = \frac{\beta}{\delta} 
\quad\text{and}\quad 
\mu = \frac{\langle d \rangle}{\langle d^2 \rangle}.
$$
A high $\lambda$ indicates a disease that is highly contagious and slow to 
%eradicate
recover from. 
On the contrary, $\mu$ describes the heterogeneity of the network. The analysis of the model shows that $\mu$ captures how much the structure of the network slows diffusion processes down: the lower the $\mu$, the more dangerous the situation is (with a physics analogy, $1/\mu$ can be though of as the \emph{conductivity} of the network with respect to the disease's diffusion process).
When $\lambda < \mu$ the disease is not endemic, and the difference between them tells us how fast it will die out.
Instead, when $\lambda \geq \mu$, the disease becomes endemic.

\begin{figure}[htbp]
	\begin{subfigure}[t]{.49\textwidth}
		\includegraphics[width=\textwidth]{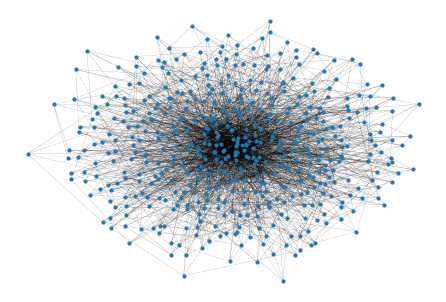}
		\caption*{Original social network}
		\includegraphics[width=\textwidth]{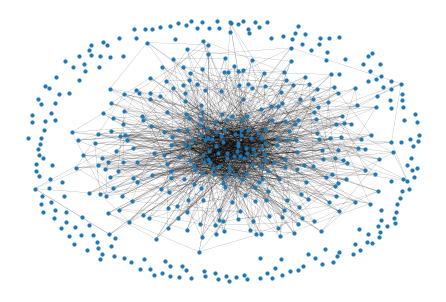}
		\caption*{$h=5$}
	\end{subfigure}
	\begin{subfigure}[t]{.49\textwidth}
		\includegraphics[width=\textwidth]{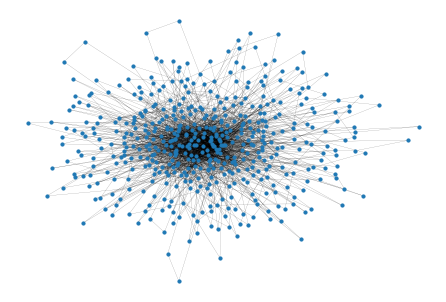}
		\caption*{$h=3$}
		\includegraphics[width=\textwidth]{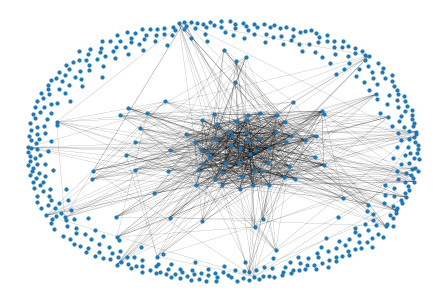}
		\caption*{$h=10$}
	\end{subfigure}
	\caption{\textbf{A social network of 500 nodes} This is a social network with degree distribution given by P(5) = P(10) = 0.4, P(20) = 0.1, P(40) = P(50) = 0.05. Different self-isolation measures are applied, depending on the number $h$ of links removed to each node.
	In this example $\mu$ falls down as $h$ increases.}
	\label{fig: networks}
\end{figure}

Social distancing policies aim at reducing the contacts among people, thus modifying the original social network in order to cut or interrupt the transmission chain of the disease, until it dies out.
However, if not everyone responds in the same way to the policy, then the resulting network may turn out to be sparser on average, but still too dense of contacts among the most active individuals. 
This can happen, for instance, if those who have more contacts are relatively less responsive to the policy indications. 
Unfortunately, then, the new smaller network might have some properties, such as a low $\mu$, that might hinder the containment of the disease or even ``help" the disease to remain endemic among those individuals who keep on being active.

Imagine, for example, that the number of people that one meets on a daily basis ranges from five to 50. 
As usually happens in the real world (see Fig \ref{fig: networks}), many people have few connections while few individuals, called \emph{hubs}, have a lot more.
Now, say that everybody is asked to cut their meetings by the same quantity (to begin with, just three contacts, which is more than half for peripheral nodes, but proportionally very little for the  hubs), so that the new \emph{degree distribution} is shifted down, but keeps the same variance.
In this case, a $\mu$ that was originally higher than $\lambda$ may actually decrease, so that a disease may remain active for more time.
If more contacts are dropped with the same uniform rule, $\mu$ may keep decreasing, up to the point that it becomes smaller than $\lambda$, and the disease remains endemic in the society, at least as long as the population is in the new network exhibits $\mu <\lambda$.
This remains true even if an additional cut completely isolates some nodes in the network, or even most of them. 
The sub--population of the remaining ones, those who had originally many links and are still very connected, will behave as an incubator for the disease because they form now a denser sub--network. 

By contrast, a policy that imposes a proportional cut their contacts to each individual always delivers an increased $\mu$ (e.g.~it doubles if the reduction is by $50\%$ for all nodes). 
% A valid intervention by the authorities should play on both the uniform scaling that reduces contacts by a constant amount (as is the effect of closing schools for students) and on targeting those individuals with many contacts (which could be obtained by closing or regulating private activities like shops and leisure meeting points).

\section*{Conclusion}

\begin{revs}In this note, we argue that a valid social-distancing\end{revs} intervention by the authorities should play on both the uniform scaling that reduces contacts by a constant amount (as is the effect of closing schools for students) and on targeting those individuals with many contacts (which could be obtained by closing or regulating private activities like shops and leisure meeting points).

% Our claim is based on a simple SIS model (see Supporting Information), but the findings are \begin{revs} not only consistent with the empirical analysis shown in Table \ref{tab:varcovid}, but also\end{revs} with those of other models that have been  recently proposed, as in the \href{https://eehh-stanford.github.io/gceid/struct.html}{examples  from the  Stanford Human Evolutionary Ecology and Health group} or the SIR models by Anderson and colleagues \cite{anderson2020will} and Koo and colleagues \cite{koo2020}.
% The same message comes from the empirical work of Chinazzi and colleagues \cite{chinazzi2020effect}, analyzing human mobility data from airline companies.

Our claim is based on a simple SIS model (see Supporting Information) \begin{revs} and consistent with the empirical analysis shown in Table \ref{tab:varcovid}. It is also in line with\end{revs} other models that have been recently proposed, as in the \href{https://eehh-stanford.github.io/gceid/struct.html}{examples  from the  Stanford Human Evolutionary Ecology and Health group} or the SIR models by Anderson and colleagues \cite{anderson2020will} and Koo and colleagues \cite{koo2020}.
The same message comes from the empirical work of Chinazzi and colleagues \cite{chinazzi2020effect}, analyzing human mobility data from airline companies.
All these works point out that restrictions are effective only if everyone fulfills the prescriptions and limits socialization.

\begin{revs}
Several issues are not dealt with here, such as mental health consequences of the imposed isolation \cite{xiang2020joint,galea2020mental,pfefferbaum2020mental} and the impact on the capacity of the health systems, because the specific focus of this work is on proposing more efficient social-distancing policies. 
To do so, we highlight the importance of distinguishing\end{revs}
%The specific focus of our approach is to distinguish 
people by their degree of socialization, and remark that if not everybody reduces drastically and proportionally their social contacts, then such measures could have an effect opposite to the one expected.

\bigskip

\bigskip

\noindent {\bf Contributors}
Both AM and PP were responsible for analyzing the model and writing the manuscript. TR was responsible for analyzing the data and writing the manuscript.

% \noindent {\bf  Declarations of interest}

% We acknowledge funding from the Italian Ministry of Education under Progetti di 
% Rilevante Interesse Nazionale (PRIN) grant 2017ELHNNJ. 

% \noindent {\bf  Acknowledgments}
% We thank Kerstin Mierke for writing assistance.

\bibliography{biblio}



\newpage

\appendix

\section{Supporting Information}

\begin{revs}

\subsection{Empirical Analysis}

In this section we describe how we collected the data used for the empirical analysis, which are of two types: data on the number of contacts and data on the number of confirmed cases and deaths.

\subsubsection{Data on number of contacts}

The data on regional average number of contacts and their variance are constructed using the public dataset provided by Belot and colleagues \cite{belot2020six}. 
This survey contains question on the number of contacts ``on a typical working day (before the outbreak of Covid-19)'' and ``on a typical day in the last 2 weeks''.
The total number of contacts of each individual has been constructed using the number of contacts for more than 15 minutes with any person (children, adult or elderly people). 
% That is, the variables \texttt{close\_workint\_more\_child}, \texttt{close\_workint\_more\_adult}, \texttt{close\_workint\_more\_elder} for the pre-outbreak period and the variables \texttt{close\_recentint\_more\_child}, \texttt{close\_recentint\_more\_adult}, \texttt{close\_recentint\_more\_elder}, referred to the contacts in the last two weeks before the interview.
% That is, the variables:
% \begin{itemize}
%     \item   \texttt{close\_workint\_more\_child}, \texttt{close\_workint\_more\_adult}, \texttt{close\_workint\_more\_elder} for the pre-outbreak period;
%     \item   \texttt{close\_recentint\_more\_child}, \texttt{close\_recentint\_more\_adult}, \texttt{close\_recentint\_more\_elder}, referred to the contacts in the last two weeks before the interview.
% \end{itemize}
The variables used are listed in Table \ref{tab:survey_variables} and some descriptive statistics are shown in Table \ref{tab:destat}.

\begin{table}[h]
    \centering
    \caption{\textbf{Variables used from survey}}
    \begin{tabular}{c|c}
    contacts before outbreak                &   contacts in last 2 weeks before interview   \\
    \hline
    \texttt{close\_workint\_more\_child}    &   \texttt{close\_recentint\_more\_child}  \\
    \texttt{close\_workint\_more\_adult}    &   \texttt{close\_recentint\_more\_adult}  \\
    \texttt{close\_workint\_more\_elder}    &   \texttt{close\_recentint\_more\_elder}
    \end{tabular}
    \label{tab:survey_variables}
\end{table}

We have dropped 3 outliers (out of 13,023 observations in the Italian, British and South Korean regions) that is individuals with more than 800 contacts, in the pre- and post-outbreak period. Notice that 800 corresponds to twenty times the 95th percentile and more than 5 times the 99th percentile of the distribution  contacts in the pre-outbreak period.

\subsubsection{Data on number of cases and deaths}

Information on number of confirmed cases and number of death has been collected from publicly available datasets. Some descriptive statistics are shown in Table \ref{tab:destat}. 
% In Figures \ref{fig: contacts_sqcontacts}-\ref{fig: sqcontacts_deaths} we plot the pairwise relations between the variables considered in the empirical analysis.

\begin{itemize}
    \item   Data for Italy and South Korea are collected as follows.
    \begin{itemize}
        \item   Information for Italian regions (confirmed cases and deaths) are provided by \emph{Protezione Civile} and available at 
     \href{https://github.com/pcm-dpc/COVID-19 }{COVID-19 Italia - Monitoraggio situazione}.
        \item   Information for South Korean regions (confirmed cases and deaths) are available at  
     \href{https://www.kaggle.com/kimjihoo/coronavirusdataset}{Data Science for COVID-19 (DS4C)}.
        \item   For both Italian and South Korean regions, the weekly number of confirmed cases and deaths (1-8 March, 13-29 April) have been computed from the corresponding daily cumulative numbers.
    \end{itemize}
    \item   Data for the United Kingdom are collected as follows.
    \begin{itemize}
        \item   Data on the number of deaths for Wales and the regions of England are provided on a weekly basis (from Friday to Friday) and available from the \href{https://www.ons.gov.uk/peoplepopulationandcommunity/birthsdeathsandmarriages/deaths/datasets/weeklyprovisionalfiguresondeathsregisteredinenglandandwales}{Office for National Statistics}.
        \item   Data on the cumulative daily number of cases in the regions of England are available at \href{https://coronavirus.data.gov.uk/downloads/csv/coronavirus-cases_latest.csv}{Coronavirus (COVID-19) in the UK}. The weekly number of cases is computed from the cumulative daily cases for the week 13-20 March and the week from April 24 to May 1, to have this measure on the same time window used by death weekly data.
        \item   Weekly data on confirmed cases for Wales are computed from daily data and are available at \href{https://public.tableau.com/profile/public.health.wales.health.protection#!/vizhome/RapidCOVID-19virology-Public/Headlinesummary}{Public Health Wales Health Protection}.
        \item   Weekly data on deaths and cases for Northern Ireland are computed from cumulative daily data and are available at
\href{https://app.powerbi.com/view?r=eyJrIjoiZGYxNjYzNmUtOTlmZS00ODAxLWE1YTEtMjA0NjZhMzlmN2JmIiwidCI6IjljOWEzMGRlLWQ4ZDctNGFhNC05NjAwLTRiZTc2MjVmZjZjNSIsImMiOjh9}{Northern Ireland Department of Health coronavirus information}.
        \item   Weekly data on deaths and cases for Scotland are computed from cumulative daily data and are available at \href{https://www.gov.scot/publications/coronavirus-covid-19-trends-in-daily-data/}{Scottish Government coronavirus information}.
    \end{itemize}
\end{itemize}

\renewcommand{\arraystretch}{1.3}
\begin{table}[h]
\centering
\small
\addtolength{\tabcolsep}{-2pt}
 \begin{threeparttable}
	\caption{\textbf{Descriptive statistics}}
	\label{tab:destat}
	\begin{tabular}{l|c|c|c}
		\hline\hline \vspace{-2 mm}\\
		&   Pre-outbreak    & Post-outbreak                      &   $\Delta$         \\
		&                   & (last 2 weeks before interview)    &   (difference post-pre) \\
		\hline
		Confirmed Cases&  299.625&  812.188& 512.563\\
		&(648.974)&(1244.210)&(1005.220)\\
		Deaths &    9.875&  212.792&  202.917\\
		&(35.830)&(328.062)&(305.180)\\
		$\langle d \rangle$&    8.270&    3.949&   -4.321\\
		&(4.991)&(2.400)&(5.006)\\
		$\langle d^2 \rangle$& 1175.600&  194.638& -980.962\\
		&(1747.951)&(361.876)&(1629.347)\\
		N. obs. & 48 & 48 &48\\
		\hline\hline
	\end{tabular}
    \begin{tablenotes}
      \small \item Mean and standard deviation (in parantheses) of the variables for the 48 regions in the sample. $\langle d\rangle$ is the average number of contacts in each region. $ \langle d^2\rangle$ is the average squared number of contacts in each region.
    \end{tablenotes}
  \end{threeparttable}
\end{table}

In Fig \ref{fig: pairwise_variables} we plot the pairwise relations among the variables used for the empirical analysis, for the countries considered (Italy, South Korea and the U.K.), where each point is a region of a country. These plots show the high variability in these variables, even if the three countries have been hit by Covid19 in the same period.

\begin{figure}
	\centering
	\caption{\bf Pairwise relations among the variables considered}
	\label{fig: pairwise_variables}
	\begin{subfigure}{0.49\textwidth} % width of left subfigure
	    \hspace{-3cm}
		\includegraphics[width=1.43\textwidth]{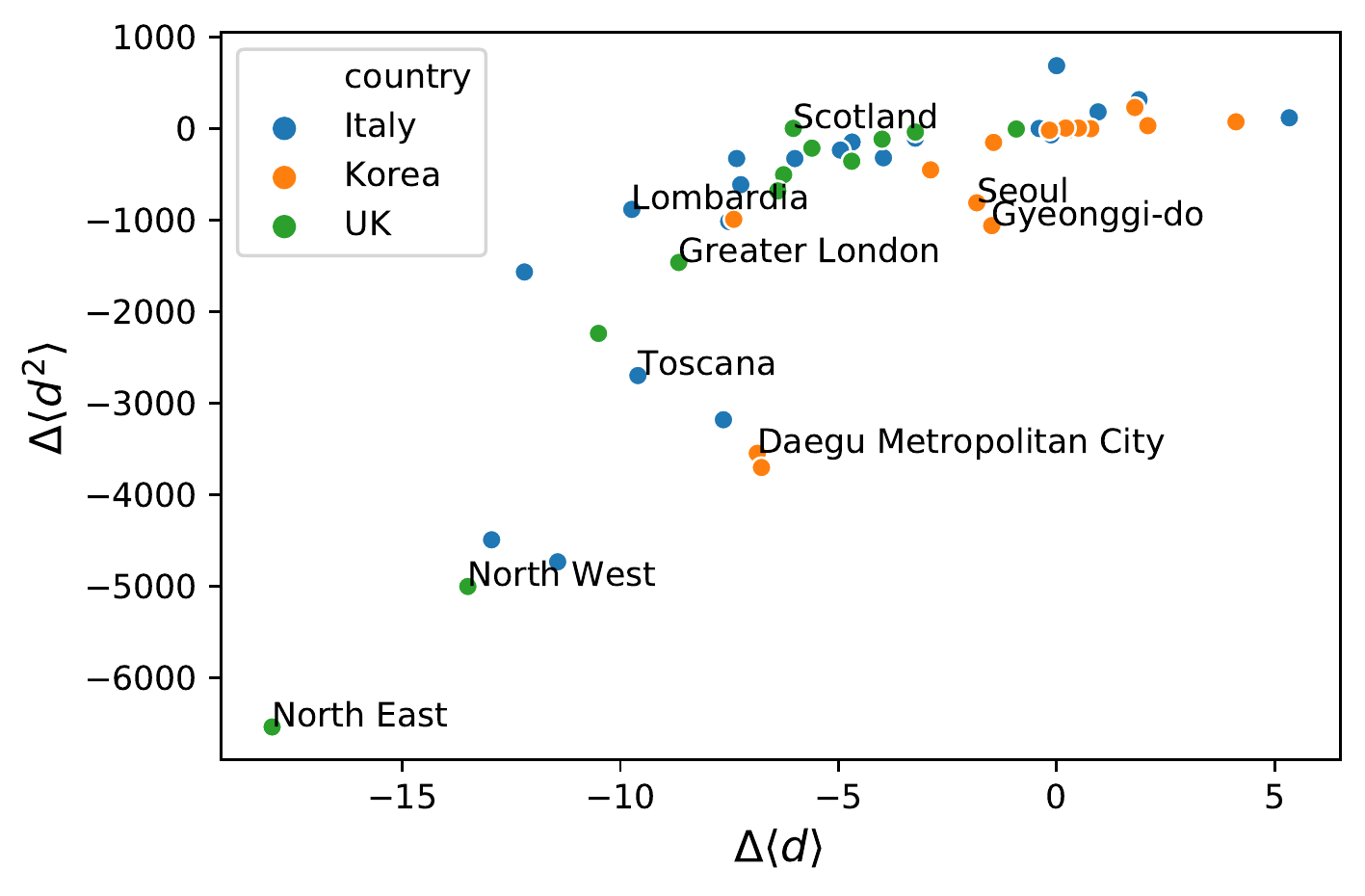}
% 		\includegraphics[width=1.4\textwidth]{img/contacts_cases.pdf}
% 		\includegraphics[width=1.4\textwidth]{img/sqcontacts_cases.pdf}
		%\caption{RNC} % subcaption
	\end{subfigure}
	%\vspace{1em} % here you can insert horizontal or vertical space
	\begin{subfigure}{0.49\textwidth} % width of right subfigure
		\includegraphics[width=1.39\textwidth]{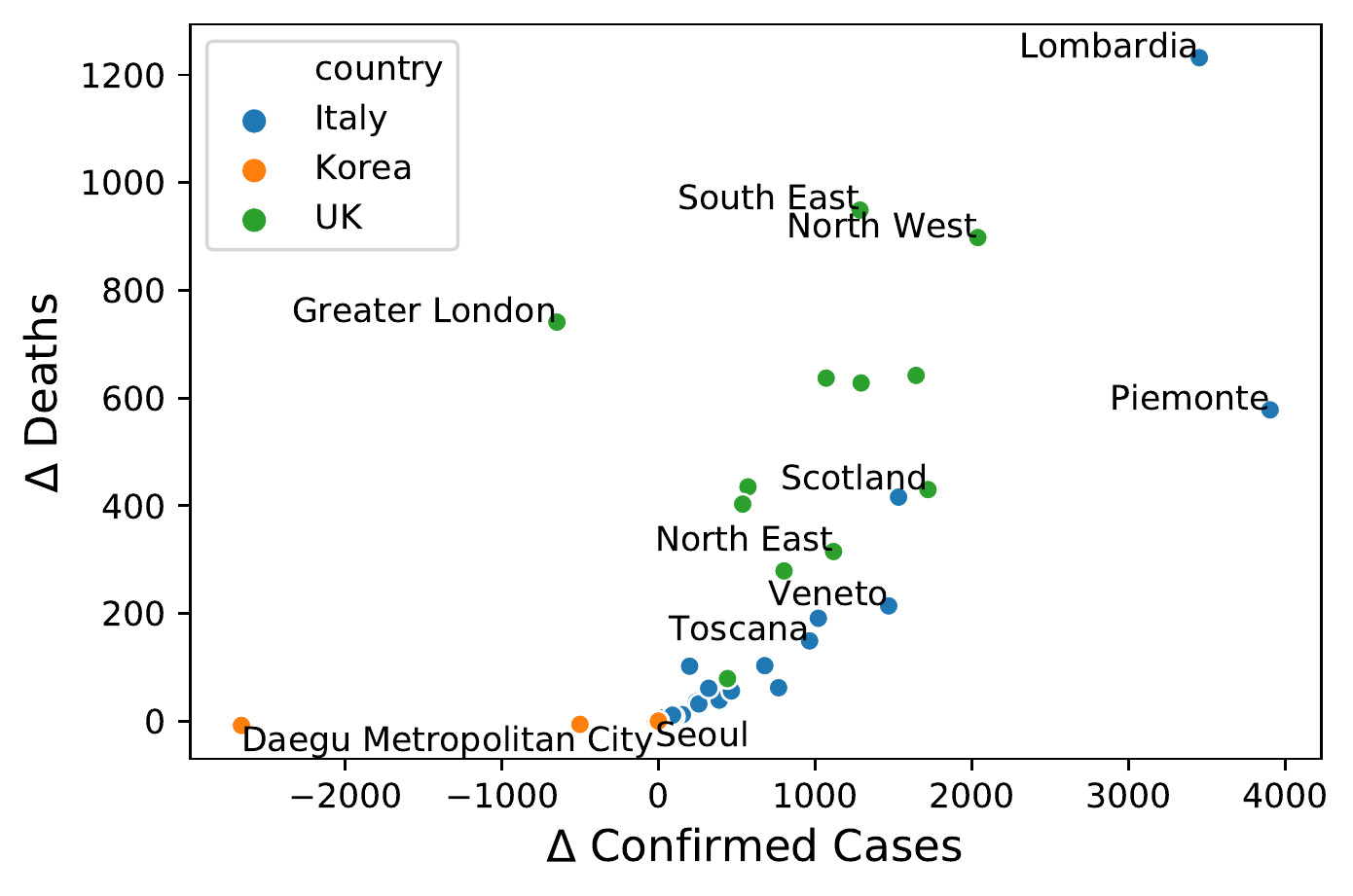}
% 		\includegraphics[width=1.4\textwidth]{img/contacts_deaths.pdf}
% 		\includegraphics[width=1.4\textwidth]{img/sqcontacts_deaths.pdf}
		%\caption{DNC} % subcaption
	\end{subfigure}
\end{figure}

\begin{figure}
	\centering
% 	\caption{}
	\begin{subfigure}{0.49\textwidth} % width of left subfigure
	    \hspace{-3cm}
 		\includegraphics[width=1.44\textwidth]{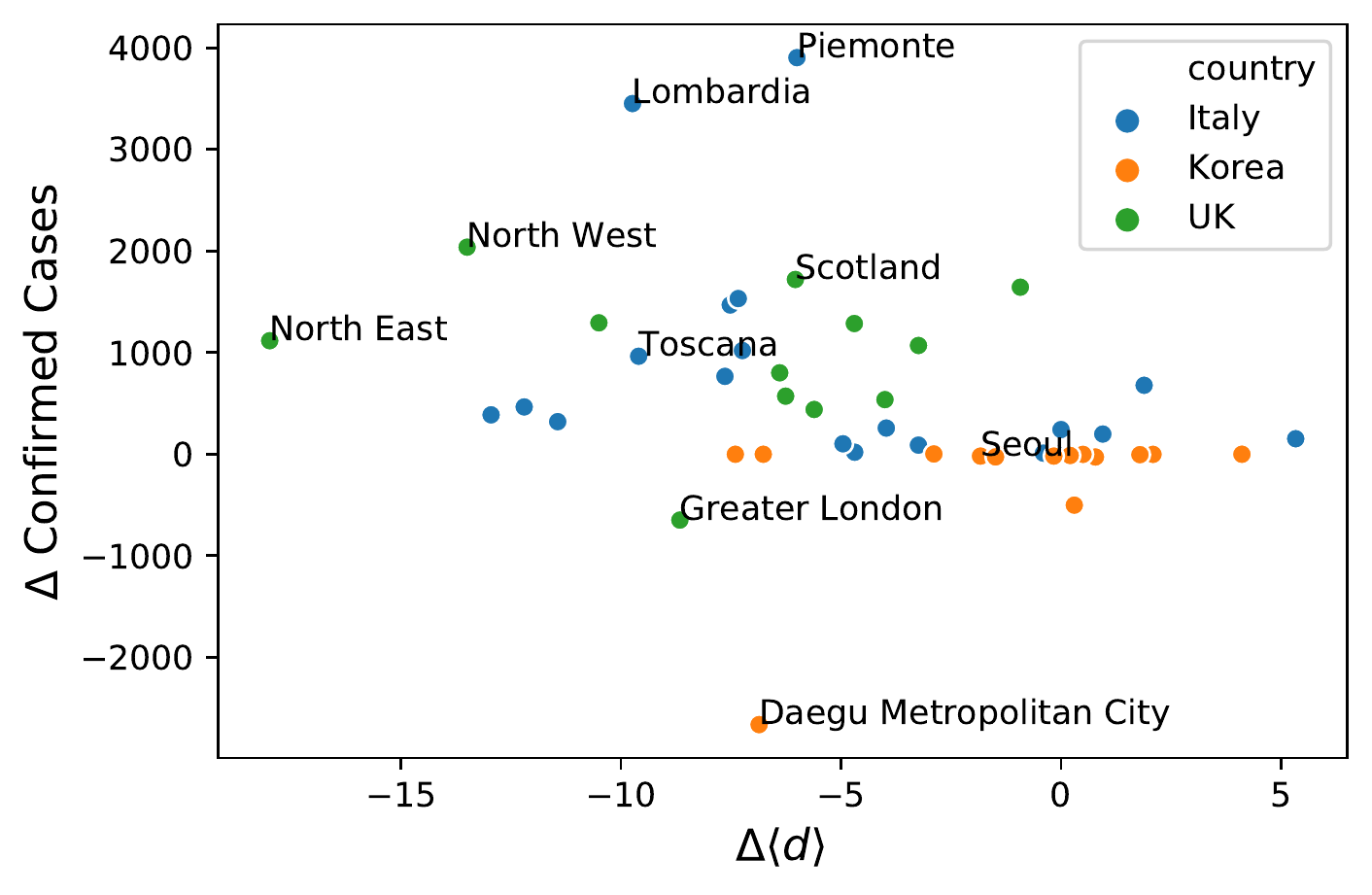}
		%\caption{RNC} % subcaption
	\end{subfigure}
	%\vspace{1em} % here you can insert horizontal or vertical space
	\begin{subfigure}{0.49\textwidth} % width of right subfigure
		\includegraphics[width=1.4\textwidth]{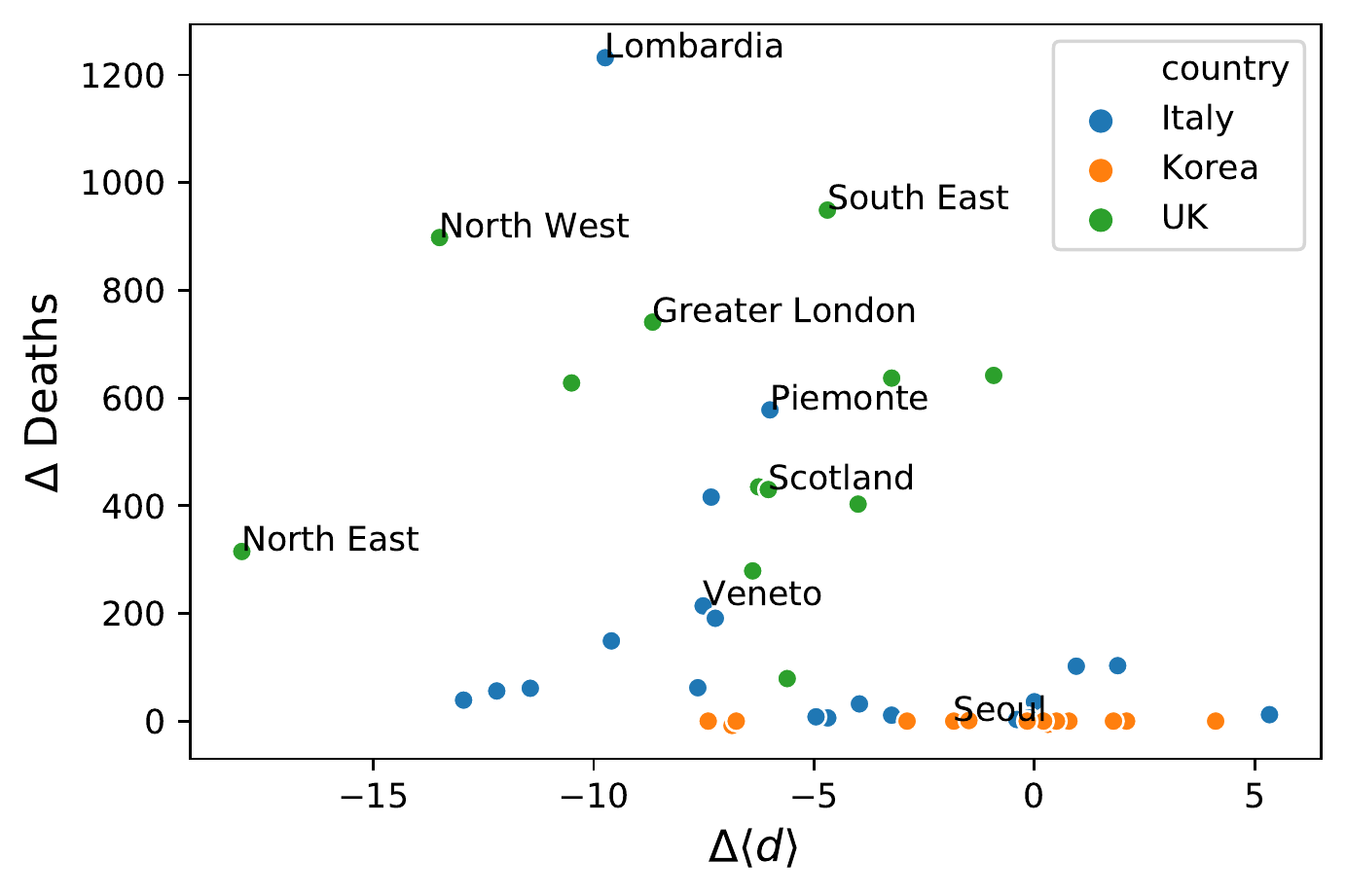}
		%\caption{DNC} % subcaption
	\end{subfigure}
\end{figure}

\begin{figure}
	\centering
% 	\caption{}
	\begin{subfigure}{0.49\textwidth} % width of left subfigure
	    \hspace{-3cm}
 		\includegraphics[width=1.48\textwidth]{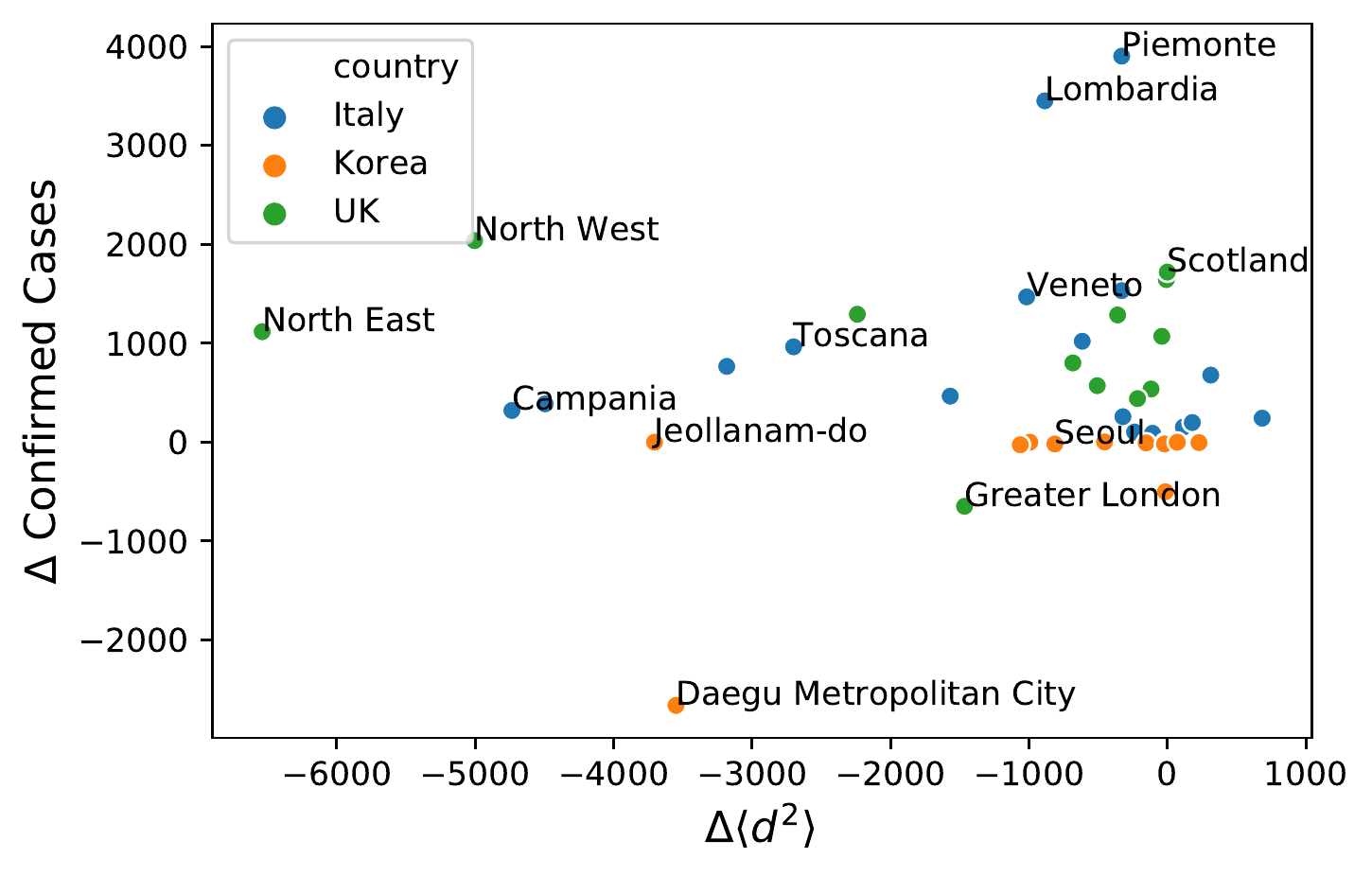}
		%\caption{RNC} % subcaption
	\end{subfigure}
	%\vspace{1em} % here you can insert horizontal or vertical space
	\begin{subfigure}{0.49\textwidth} % width of right subfigure
	   % \hspace{1mm}
 		\includegraphics[width=1.43\textwidth]{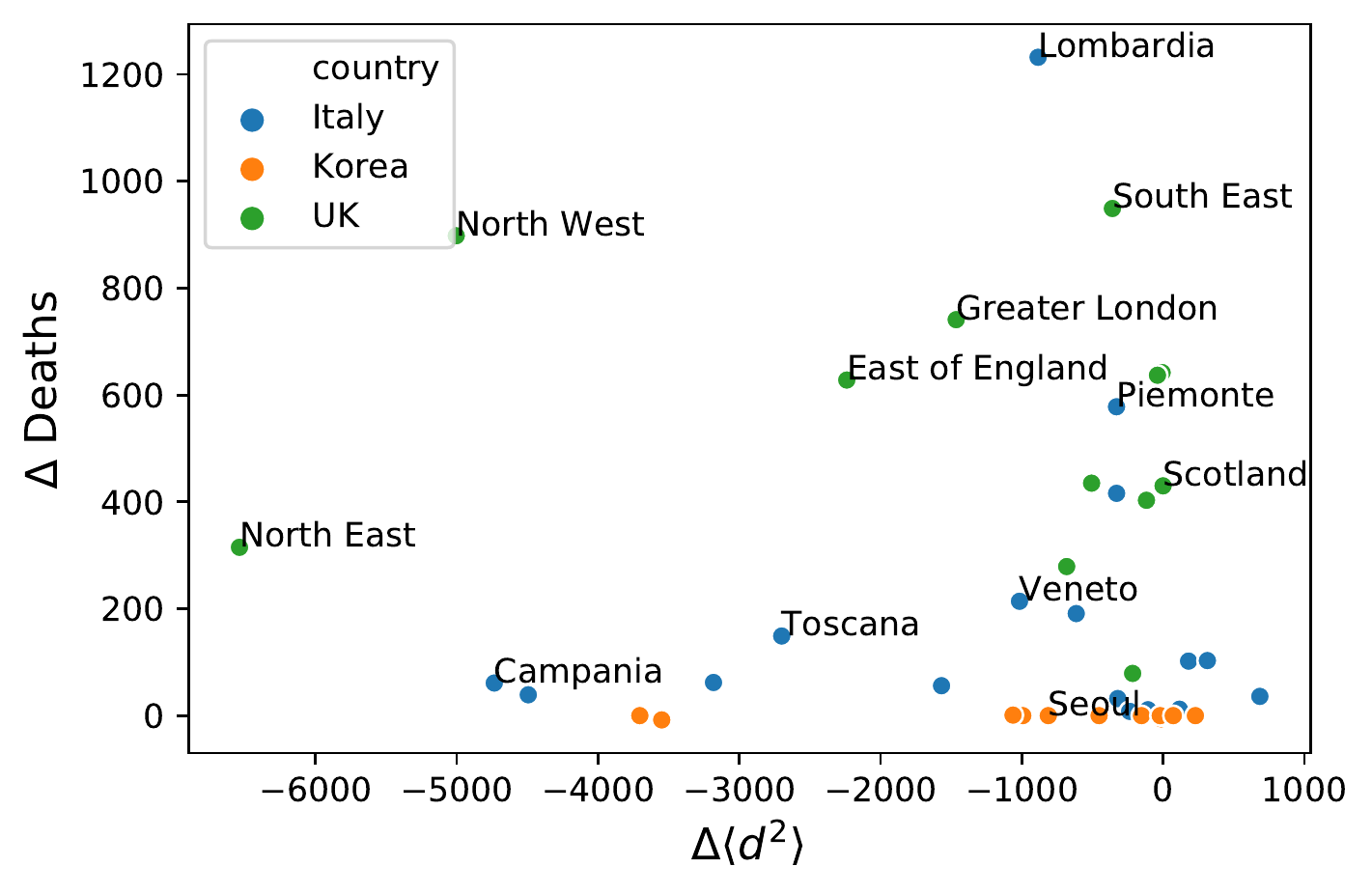}
		%\caption{DNC} % subcaption
	\end{subfigure}
\end{figure}

\subsubsection{Robustness check}

In Table \ref{tab:nlcovid} we investigate the presence on a non-linear relation between the the number of cases and deaths and the average  number of contacts. The inclusion of the square of the average number of contact does not alter the qualitative impact of $\Delta\langle d ^2\rangle$ nor its statistical significance.

\renewcommand{\arraystretch}{1.3}
\begin{table}[h]
\centering
\small
\addtolength{\tabcolsep}{5pt}
 \begin{threeparttable}
	\caption{\bf Variations in cases and deaths }
	\label{tab:nlcovid}
	\begin{tabular}{l|cc|cc}
		\hline\hline \vspace{-2 mm}\\
           &    \multicolumn{2}{c}{$\Delta$  Confirmed Cases}  &   \multicolumn{2}{c}{$\Delta$ Deaths}  \\
                      &        (1)   &     (2)   &        (3)   &        (4)\\
                      \hline
 $\Delta \langle d \rangle$    &     -67.945   &     -67.599   &     -27.361   &     -27.275   \\
            &    (64.513)   &    (61.189)   &    (19.401)   &    (19.083)   \\
  $\Delta\langle d \rangle^2$&      -0.067   &      -7.971*  &       0.094   &      -1.862   \\
            &     (3.113)   &     (4.429)   &     (1.148)   &     (1.328)   \\
  $\Delta\langle d ^2\rangle$ &               &       0.588** &               &       0.146***\\
            &               &     (0.219)   &               &     (0.048)   \\
Constant       &     214.174** &     227.193** &      91.448***&      94.671***\\
            &    (89.785)   &    (98.383)   &    (31.568)   &    (32.896)   \\
N. obs.           &      48   &      48   &      48   &      48   \\
 Adj. $R^2$       &       0.080   &       0.297   &       0.139   &       0.278   \\
		\hline\hline	
\end{tabular}
\begin{tablenotes}
      \small
      \item Note: In columns 1 to 2 the dependent variable is the variation in the numbers of confirmed cases in a region. In columns 3 to 4 the dependent variable is the variation in the number of deaths in each region. $\Delta\langle d\rangle$ is the variation in the average number of contacts in each region. $\Delta \langle d^2\rangle$ is the variation in the average squared number of contacts in each region. $\Delta \langle d\rangle^2$ is the variation in the square of average number of contacts in each region. *** significant at 1\%, ** significant at 5\%, * significant at 10\%.
    \end{tablenotes}
  \end{threeparttable}
\end{table}

\end{revs}

\subsection{The model}

Consider a society formed by a large number of individuals who interact by meeting others at random and where each individual can alternate between being susceptible or infected to a disease which transmits via social contacts.
More in detail, we first consider a degree-based random mixing model with an infinite number of agents, often thought as an ``approximation'' of a large social network. Then, we consider a tractable first approximation of a susceptible-infected-susceptible (SIS) model based on a simple linear form, standard in the literature \cite{pastor2001epidemic,jackson2010social}.
Although we are not the first ones to consider this model in a context of quarantine, all previous works have focused on different issues and did not consider the possible negative effects of a quarantine that is not homogeneous \cite{lagorio2011quarantine,kang2017dynamics,li2019dynamic}.
Technically, we adopt a SIS model because its ergodic nature  delivers neat analytical results. Moreover, it is still not clear to scientists whether Covid-19 can affect more than once the same person. 
Cases of multiple infection in the same person have been reported \href{https://www.reuters.com/article/us-china-health-japan/japanese-woman-confirmed-as-coronavirus-case-for-second-time-weeks-after-initial-recovery-idUSKCN20L0BI}{by the end of February in China and Japan}.

Consider a network with degree distribution $P(d)$, i.e. where the degree of a node $i$ is $d_i$ and $P(d)$ is the fraction of individuals with degree $d$. The probability of meeting an agent of degree $d$ is $P(d) d/\langle d \rangle$. Let $\rho(d)$ be the fraction of individuals of degree $d$ who are currently infected, so that the probability of meeting an infected agent of degree $d$ is
$$
\rho(d) \frac{P(d) d}{\langle d \rangle}.
$$
Overall, the probability of meeting an infected individual is
\begin{equation}
\label{eq: theta}
\theta = \sum_d \rho(d) \frac{P(d) d}{\langle d \rangle},
\end{equation}
while the average infection rate in the population is $\rho = \sum_d \rho(d) P(d)$.

The mechanism of the disease transmission is as follows. The chance that a given individual of degree $d$ becomes infected in a given period when faced with a probability $\theta$ that any given meeting is with an infected individual is
$$
\beta \theta d,
$$
where $\beta \in (0,1)$ is a parameter describing the rate of transmission of the infection in a given period. The probability that an infected individual recovers (and becomes again susceptible) in a given period is $\delta \in (0,1)$.

With a mean-field approach, one can compute the expected change of $\rho(d)$ over time, for all $d$
\begin{equation}
\label{eq: mean_field_equation}
\frac{\text{d} \rho(d)}{\text{d}t} = (1 - \rho(d)) \beta \theta d - \rho(d) \delta,
\end{equation}
where the first term describes the inflow of susceptibles becoming infected and the second term describing the outflow, i.e. infected who recover.

The steady-state of the system is such that $\text{d} \rho(d)/\text{d}t = 0$. Solving this equation yields
\begin{equation}
\label{eq: rho_steady_state}
\rho(d) = \frac{\lambda \theta d}{\lambda \theta d + 1},
\end{equation}
where $\lambda := \beta/\delta$. Plugging Eq. \eqref{eq: rho_steady_state} into Eq. \eqref{eq: theta} gives the condition
\begin{equation}
\label{eq: condition_theta_and_H}
\theta = \underbrace{\frac{1}{\langle d \rangle} \sum_d \frac{\lambda \theta d^2 P(d)}{\lambda \theta d + 1}}_{=: H(\theta)}.
\end{equation}
The function $H(\theta)$ keeps track of how many individuals would become infected starting from a level $\theta$. Steady states of the system are fixed points such that $H(\theta) = \theta$ and Eq. \eqref{eq: condition_theta_and_H} has always solution $\theta = 0$, but can also have other solutions. Since $H(0) = 0$ and $H(\theta)$ is increasing and strictly concave in $\theta$, then it turns out that in order to have a (unique) positive steady state it must be that $H'(\theta) > 1$. Since $H'(\theta) = \lambda \langle d^2 \rangle/\langle d \rangle$, then the condition for an endemic equilibrium to exist is (and corresponding also to a positive average infection rate in the population, $\rho > 0$)
\begin{equation}
\label{eq: esistence_endemic_equilibrium}
\lambda > \underbrace{\frac{\langle d \rangle}{\langle d^2 \rangle}}_{=: \mu}.
\end{equation}
This condition means that the infection-to-recovery ratio has to be high enough relative to average degree divided by second moment (roughly variance of degree distribution). 
Intuitively, this shows that high degree nodes are more prone to infection and, since they have many meeting, also serve as conduits for infection. In general, a social network with high variance in the degree distribution is such that there are many of such high degree nodes.

\subsubsection{Endemic Disease from Self-isolation}

From Eq. \eqref{eq: esistence_endemic_equilibrium}, we have that if $\mu = \langle d \rangle / \langle d^2 \rangle$ decreases, then the epidemics can become endemic. 
This can happen, for example, if during a self-isolation period only the nodes with low degree reduce drastically their contacts.

In general, consider the situation in which all nodes decrease their contacts by a common discrete number $h$, obtaining a new re-scaled degree distribution $\hat{d} = d - h$. 
The mean degree becomes $\langle \hat d \rangle = \langle d \rangle - h$, but the variance of the degree distribution $\langle \hat d^2 \rangle - \langle \hat d \rangle^2 = \langle d^2 \rangle - \langle  d \rangle^2$ remains unchanged. However, this new distribution is such that the threshold $\langle \hat d \rangle / \langle \hat d^2 \rangle$ in Eq. \eqref{eq: esistence_endemic_equilibrium} is:
$$
\frac{\langle d - h \rangle}{\langle (d - h)^2 \rangle} = \underbrace{\frac{\langle d \rangle - h}{\langle d^2 \rangle - 2 h \langle d \rangle + h^2}}_{=: \mu(h)}.
$$
Since 
$$
\frac{\text{d}\mu}{\text{d}h}\Big|_{h=0} = \frac{-\langle d^2 \rangle + 2 \langle d \rangle^2}{\langle d^2 \rangle^2},
$$ 
then it is negative when $\langle d^2 \rangle > 2 \langle d \rangle^2$, which holds if the standard deviation is high enough. 
For $h$ small, this marginal effect remains negative which indicates that $\mu(h)$ decreases. Specifically, as $h$ increases then $\mu(h)$ decreases as long as $h$ does not exceed
$$
2 \langle d \rangle - \frac{\langle d^2 \rangle}{\langle d \rangle}.
$$ 
This implies that if the cut to links imposed by the self-isolation policy is too weak, i.e. $h$ is too small, then the threshold for the existence of the endemic equilibrium decreases. Thus, a disease that was not endemic may instead become endemic.

\subsubsection{Speed of Recovery to Disease-free Equilibrium}

From Eq. \eqref{eq: mean_field_equation} we can compute the Jacobian $J$ when the disease is not endemic.
That is, when $\rho(d) \rightarrow 0$ for all $d$, and also $\theta \rightarrow 0$.
Deriving Eq. \eqref{eq: mean_field_equation} with $\rho = 0$ and $\theta = 0$ yields
$$
J_{k \ell} = 
\begin{cases}
\frac{\beta}{\langle d \rangle} k^2 P(k) - \delta,  & \text{if } k = \ell,  \\
\frac{\beta}{\langle d \rangle} k \ell P(\ell),     & \text{if } k \neq \ell,
\end{cases}
$$
which can be written as the $(D \times D)$-matrix $J = (J_{k \ell})_{k,\ell = 1, \dots, D}$ of the form (where $D$ is the maximum degree in the network)
$$
 J = 
\frac{\beta}{\langle d \rangle}
\begin{pmatrix}
1       \\
\vdots  \\
k       \\
\vdots  \\
D
\end{pmatrix}
\begin{pmatrix}
1 P(1) \cdots \ell P(\ell) \cdots D P(D) 
\end{pmatrix}
- \delta  I,
$$
where in the first term there is the matrix multiplication between two vectors and $ I$ is the identity matrix. 

In general, consider a matrix $ A :=  u  v' - \delta  I$. Then, its eigenvalues are $-\delta$ and $ v'  u - \delta$. The corresponding eigenvectors are, respectively, all vectors orthogonal to $ v$ and $ u$ itself. In our case, then, the only eigenvalues of $ J$ are
$$
e_1 = - \delta 
\qquad\text{and}\qquad
e_2 = \beta \frac{\langle d^2 \rangle}{\langle d \rangle} - \delta \equiv \frac{\beta}{\mu} - \delta.
$$
While $e_1 = - \delta$ is independent of the network and always negative, $e_2$, which is proportional to the difference $1/\mu - 1/\lambda$, is negative if and only if $\lambda < \mu$. From Eq. \eqref{eq: esistence_endemic_equilibrium}, this occurs exactly when the only equilibrium is the disease-free equilibrium and it is asymptotically stable.

Moreover, from the policy perspective, in this case the speed of convergence to the disease-free equilibrium is determined by 
$$
|e_2| = \delta - \frac{\beta}{\mu}.
$$
This implies that as $\mu = \langle d \rangle/\langle d^2 \rangle$ increases, so does $|e_2|$ and the speed of convergence to the disease-free equilibrium increases as well.
Conversely, as $\mu$ decreases, so does the speed of convergence, up to the point where $\mu$ goes below the threshold $\lambda$ in Eq. \eqref{eq: esistence_endemic_equilibrium}, which is when the equilibrium becomes endemic.

\end{document}